\documentclass[twocolumn,twoside,slac_two]{revtex4}
\usepackage{graphicx}
\usepackage{a4p}
\usepackage{fancyhdr}
\usepackage{atlasphysics}
\usepackage{multirow}
\usepackage{textcomp}
\usepackage{booktabs}
\usepackage{bm}
\usepackage{natbib}

\newcommand{\mm}{$\mu^{+^{}}\mu^{-^{}}$}
\newcommand{\ep}{$e^{+^{}}e^{-^{}}$}
\newcommand{\x}{$\tilde{\chi}^{0}_{1}$}

\newcommand{\zgmet}{$Z\gamma\ +\not\!\!E_{T} \ $}

\newcommand{\xxzggg}{$\tilde{\chi}^{0}_{1}\tilde{\chi}^{0}_{1}\ra Z(\ell^{+}\ell^{-})\gamma\tilde{G}\tilde{G}$}
\newcommand{\xxzmmggg}{$\tilde{\chi}^{0}_{1}\tilde{\chi}^{0}_{1} \ra Z(\ra \mu^{+}\mu^{-})\gamma\tilde{G}\tilde{G}$}
\newcommand{\xxzeeggg}{$\tilde{\chi}^{0}_{1}\tilde{\chi}^{0}_{1}\ra Z(\ra e^{+}e^{-})\gamma\tilde{G}\tilde{G}$}

\newcommand{\zll}{$Z(\ra \ell^{+}\ell^{-}) \ $}
\newcommand{\zmmX}{$Z(\ra \mu^{+^{}}\mu^{-^{}}) + X \ $}
\newcommand{\zeeX}{$Z(\ra e^{+^{}}e^{-^{}}) + X \ $}
\newcommand{\zllX}{$Z(\ra \ell^{+}\ell^{-}) + X \ $}
\newcommand{\zeeg}{$Z($\ep$)\gamma \ $}
\newcommand{\zmmg}{$Z($\mm$)\gamma \ $}
\newcommand{\zmmgmet}{$Z(\mu^{+}\mu^{-})\gamma\ +\not\!\!E_{T} \ $}
\newcommand{\zeegmet}{$Z(e^{+}e^{-})\gamma\ +\not\!\!E_{T} \ $}

\begin{document}

\title{Discovery Potential for GMSB Supersymmetry in ATLAS using the $Z\gamma + \not\!\!E_{{T}}$ Final State at a center of mass energy of $\sqrt{s}$=10 \TeV}

%

\author{D.~Harper for ATLAS collaboration}
\affiliation{Department of Physics, University of Michigan, Ann Arbor, Michigan, USA}
\begin{abstract}
We have studied the sensitivity of the ATLAS detector for supersymmetric neutralino signals in the $Z\gamma \ +\not\!\!E_{{T}}$ final state in a GMSB model in which the Higgsino-like neutralino is NLSP. This study considers the reaction of $pp \ra \tilde{\chi}^{0}_{1} \tilde{\chi}^{0}_{1} \ra Z(\ra \ell^{+}\ell^{-})\gamma \tilde{G}\tilde{G}$, where $\ell = \mu/e$, at a center of mass energy of $\sqrt{s}$=10 \TeV using fully simulated ATLAS Monte Carlo events both for the signal and background. Based on the GMSB Model Line E predictions, we expect that, for an integrated luminosity of less then 3 fb$^{-1}$, ATLAS could detect the GMSB signal from the Higgsino-like neutralino that has a mass of 134.7 \GeV in the $Z\gamma \ +\not\!\!E_{{T}}$ final state with a significance of 5$\sigma$, assuming 20$\%$ systematic uncertainty.
\end{abstract}

\maketitle

\thispagestyle{fancy}


\section{Introduction}
One of the major motivations of the LHC experiments is to search for physics beyond the Standard Model (SM). Supersymmetry (SUSY) is a favored candidate in this search. However, SUSY must be broken so that the masses of super particles exceed those of their SM partners. The minimal supersymmetric extension of the standard model (MSSM) does not incorporate SUSY breaking and has 124 free parameters. SUSY breaking can be achieved by introducing a hidden sector. The mediation of the breaking from the hidden sector to the observable MSSM sector is enabled by different mechanisms such as mSUGRA, gauge-mediated SUSY breaking (GMSB) and AMSB through gravity, gauge or anomalous interactions. Through this mediation, the number of free parameters is significantly reduced. 

We have investigated neutralino pair production at the LHC at a center of mass energy of $\sqrt{s}$ = 10 \TeV within the framework of a GMSB scenario~\cite{theory}. A distinctive phenomenological feature of the GMSB model is the presence of neutral dibosons ($Z$ or $\gamma$) plus missing transverse energy ($\not\!\!E_{{T}}$) in the final state. 

This note reports on the sensitivity of the ATLAS detector at $\sqrt{s}$ = 10 \TeV for the process $pp \ra \tilde{\chi}^{0}_{1} \tilde{\chi}^{0}_{1} \ra Z(\ra \ell^{+}\ell^{-})\gamma \tilde{G}\tilde{G}$, where $\ell = \mu/e$. \x \ in this process is a Higgsino-like neutralino, which is the next-to-lightest supersymmetric particle (NLSP) in the model considered here~\cite{Baer}. It decays to a neutral vector boson ($Z$ or $\gamma$) and a gravitino ($\tilde{G}$), the lightest supersymmetric particle (LSP), which escapes detection resulting in large $\not\!\!E_{{T}}$.

\subsection{Theoretical overview}

In GMSB models, SUSY, which is broken at a \TeV scale in a hidden sector, is propagated down to the MSSM observable sector via new chiral supermultiplets, called messengers, that couple indirectly to the MSSM particles through the ordinary $SU(3)_{C} \times SU(2)_{L} \times U(1)_{Y}$ gauge boson and gaugino interactions. The main advantage of GMSB models is the automatic creation of identical soft SUSY breaking masses for scalars with the same gauge quantum numbers but different flavors. Therefore, there are no problems with flavor changing neutral currents (FCNC) or CP violation constraints.

Within the minimal version of the GMSB framework (mGMSB), the couplings, branching ratios, decay widths and sparticles masses are determined by the parameters:
\begin{equation}
\Lambda, M, N_{5}, \tan(\beta), sign(\mu), C_{grav}.
\end{equation}
$\Lambda$ is the SUSY breaking energy scale, $\Lambda = F/M$, where $F$ is a vacuum expectation value of an auxiliary field that determines the magnitude of supersymmetry breaking in the vacuum state. $M$ is the size of the messenger mass scale, $M > \Lambda$. For electroweak scale superpartners, $\Lambda$ is $\sim 100\TeV/\sqrt N_{5}$. $N_{5}$ is the number of generations of messenger fields, $\tan(\beta)$ is the ratio of the MSSM Higgs vacuum expectation values ($\langle H^{0}_{u}\rangle/\langle H^{0}_{d}\rangle$), $sign(\mu)$ is the sign of the Higgs sector mixing parameter and $C_{grav}$ is the ratio of the messenger sector SUSY breaking order parameter to the intrinsic SUSY breaking order parameter, which controls the coupling to the gravitino.

In GMSB, $\tilde{G}$ is the LSP (usually, \mbox{M($\tilde{G})$ $<<$ 1 \GeV}). The $\tilde{G}$ has a derivative coupling to each particle and its superpartner with an interaction strength that is inversely proportional to $\sqrt{F}$. Because of this coupling, the next-to-lightest superpartner is unstable and decays to its lighter partner through $\tilde{G}$ emission. As a result, the nature of the NLSP defines the phenomenology of the GMSB model. Depending on the region in parameter space, either the neutralino($\tilde{\chi}^{0}_{1}$) or the $\tilde{\tau}$ arise as the NLSP. The NLSP decay length, which depends on $\sqrt{F}$, can be divided into three ranges: (1) the NLSP decays promptly, (2) the NLSP decays inside the detector away from the collision point ($\sqrt{F} \lesssim 10^6 \GeV$),  and (3) the NLSP decays outside the detector ($\sqrt{F} >10^6 \GeV$). We consider only the case where the NLSP is the \x, which decays promptly, corresponding to $C_{grav} = 1$.

$\tilde{\chi}^{0}_{1}$ is a mixture of gaugino ($\tilde{B}$, $\tilde{W^{0}}$) and Higgsino ($\tilde{H^{0}_{u}}$, $\tilde{H^{0}_{d}}$) eigenstates, and therefore \x \ decays to a $\gamma$, $Z$, or Higgs (h). If \x \ is gaugino-like, it decays mostly to $\gamma\tilde{G}$, leading to the $\gamma\gamma \ + \not\!\!E_{T}$ signature (with R-parity conservation, supersymmetric particles are produced in pairs). If the \x \ is Higgsino-like, it decays to $h \tilde{G}$. In addition, since the longitudinal component of the $Z$ boson mixes with the Goldstone mode of the Higgs field, a Higgsino-like neutralino also decays to $Z\tilde{G}$. Because of a strong phase space suppression of the h and $Z$ final states, decay to a $\gamma$ can also be important for Higgsino-like neutralinos, which are not very much heavier than the $Z$ boson. Consequently, a pair of Higgsino-like neutralinos produced in a collider can give rise to the diboson final states \mbox{($\gamma\gamma,hh,h\gamma, hZ,Z\gamma,ZZ) \ +\not\!\!E_{{T}}$}~\cite{f_Matchev}, \cite{ferm}.
We consider the possibility of a Higgsino-like $\tilde{\chi}^{0}_{1}$ NLSP, where one neutralino decays to $\gamma \tilde{G}$ and the other decays to $Z\tilde{G}$, leading to the $\mbox{\zgmet}$ signature.

\subsection{$\mbox{\zgmet}$ signature}
To study the $\mbox{\zgmet}$ signature in the GMSB framework, we generated a sample of $pp \ra \tilde{\chi}^{0}_{1} \tilde{\chi}^{0}_{1} \ra Z(\ell^{+}\ell^{-})\gamma \tilde{G}\tilde{G}$ events, where $\ell = \mu/e$, using a (non-minimal GMSB) Higgsino-like neutralino model (model line E), described in~\cite{Baer}. This model is characterized by the following parameters:   
\begin{equation}
M/\Lambda = 3, N_{5} = 2, \tan(\beta) = 3, \mu = 0.75M_{1},
\end{equation}
where $M_{1}$ is the gaugino mass and yields 225.4 \GeV. Unlike in the mGMSB, where the absolute value of $\mu$ is set by the electroweak symmetry breaking (EWSB) radiative conditions as shown in equation [3], $\mu$ is set to 0.75$M_{1}$ in this approach, which gives rise to a large Higgsino fraction in the neutralino. 
\begin{equation}\mu^{2} = - \frac{m^{2}_{Z}}{2} + \frac{m^{2}_{H_{u}} - m^{2}_{H_{d}}\tan^{2}(\beta)}{\tan^{2}(\beta) - 1}
\end{equation}

\begin{figure*}[tb]
\centering
\begin{tabular}{cc}
\includegraphics[width=85.5mm]{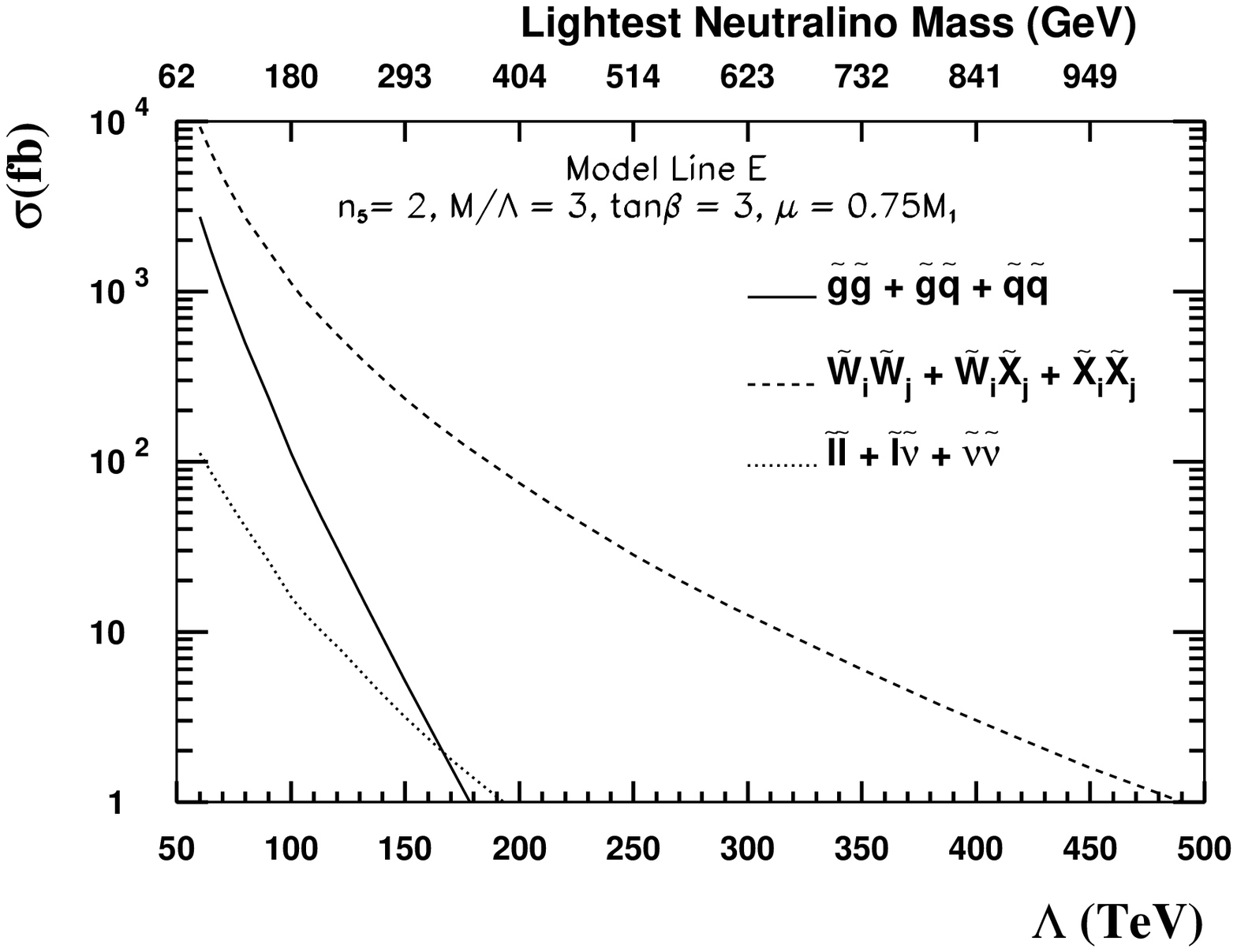} &
\includegraphics[width=80mm]{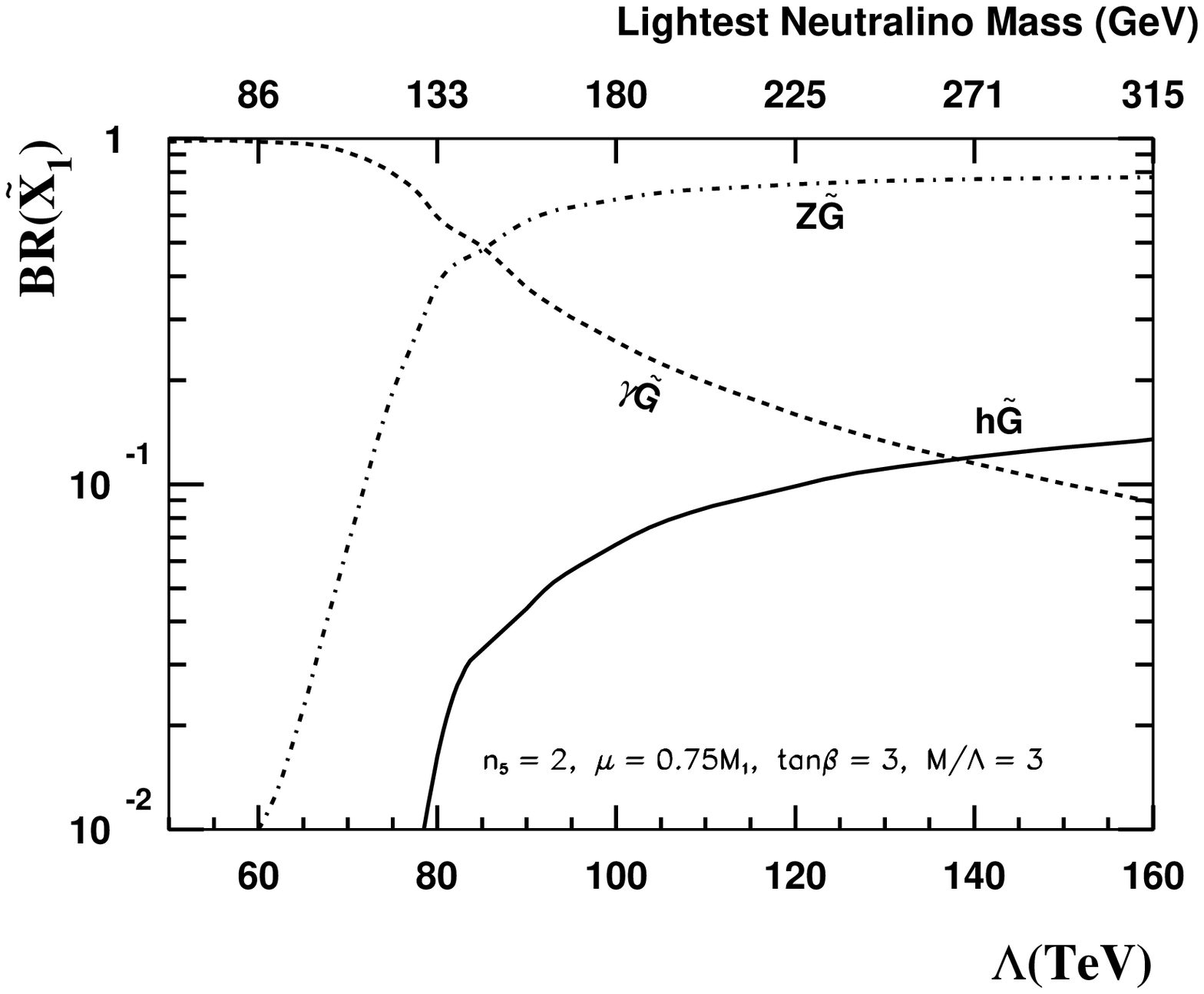} \\
\end{tabular}
\caption{Cross-sections for various sparticle production processes in model line E (left). The branching fractions for various decays of the \x \ NLSP (right) in model line E versus the parameter $\Lambda$. The scale on top shows the \x \ mass.} \label{fig:t_br}
\end{figure*}
Figure~\ref{fig:t_br}, taken from~\cite{Baer}, shows the production cross-sections for sparticles(left plot) and branching fractions of \x (right plot) in model line E, as a function of the most influential parameter, $\Lambda$. 
The dominant source of sparticles in this model is production of neutralino pairs, chargino pairs ($\tilde{\chi}^{\pm}_{i}$, i = 1,2), and chargino-neutralino pairs. Charginos and heavier neutralinos decay to \x \ through cascade decays. Gluinos and squarks are relatively heavy; therefore, their production becomes suppressed when $\Lambda$ is increased~\cite{Baer}. The branching ratio Br($\tilde{\chi}^{0}_{1} \ra Z\tilde{G}$) becomes dominant for $\Lambda >$ 80 TeV (M(\x)$\sim$134 \GeV); we thus choose this value of $\Lambda$ for the current study. 
This choice gives a production cross-section of the decay \xxzggg, where $\ell = \mu/e$, of about 56.2 fb. The cross-section is calculated as follows: 
$2\times \sigma(pp \ra \tilde{\chi}^{0}_{1} \tilde{\chi}^{0}_{1}$) $\times$ Br($\tilde{\chi}^{0}_{1} \ra Z \tilde{G}$) $\times$ Br($\tilde{\chi}^{0}_{1} \ra \gamma \tilde{G}$) $\times$ Br($Z \ra \ell^{+}\ell^{-})$, where $\sigma(pp \ra \tilde{\chi}^{0}_{1} \tilde{\chi}^{0}_{1}$) = 1880 fb, Br($\tilde{\chi}^{0}_{1} \ra \gamma \tilde{G}) = 0.55$, Br($\tilde{\chi}^{0}_{1} \ra Z \tilde{G}) = 0.4$, and Br($Z \ra \ell^{+}\ell^{-}) =  0.067$~\cite{pdg}. The SUSY particle mass spectrum for model line E with $\Lambda =$ 80 TeV is listed in Table~\ref{tab:par}. 

\begin{table}[h]
\begin{center}
\caption{SUSY particle mass spectrum for the GMSB model characterized by the following parameters: $\Lambda=80 \TeV,\ M = 240 \TeV,\ N_{5} = 2,\ \tan(\beta) = 3,\ \mu = 169 \GeV, M(t) = 172.5 \GeV$.}
\begin{tabular}{|l|c|c|}
\hline
\textbf{Names} & \textbf{Mass Eigenstates} &  \textbf{M (\GeV)} \\
\hline
\multirow{12}{*}{squarks} 
& $\tilde{u}_{L}$ & 1272.9  \\
& $\tilde{u}_{R}$ & 1220.6  \\
& $\tilde{d}_{L}$ & 1272.9  \\
& $\tilde{d}_{R}$ & 1215.6  \\
& $\tilde{s}_{L}$ & 1272.9  \\
& $\tilde{s}_{R}$ & 1215.6  \\
& $\tilde{c}_{L}$ & 1272.9  \\
& $\tilde{c}_{R}$ & 1220.6  \\
& $\tilde{t}_{1}$ & 1119.1  \\
& $\tilde{t}_{2}$ & 1239.2  \\
& $\tilde{b}_{1}$ & 1216.9  \\
& $\tilde{b}_{2}$ & 1228.4  \\
\hline
\multirow{6}{*}{sleptons} 
& $\tilde{e}_{L}$       &  408.9\\
& $\tilde{e}_{R}$       &  196.1\\
& $\tilde{\mu}_{L}$     &  408.9\\
& $\tilde{\mu}_{R}$     &  196.1\\
& $\tilde{\tau}_{1}$    &  198.9\\
& $\tilde{\tau}_{2}$    &  407.9\\
\hline
\multirow{4}{*}{neutralinos} 
& $\tilde{\chi}^{0}_{1}$     &  134.7  \\
& $\tilde{\chi}^{0}_{2}$     &  172.3  \\
& $\tilde{\chi}^{0}_{3}$     &  241.3  \\
& $\tilde{\chi}^{0}_{4}$     &  443.3  \\
\hline
\multirow{2}{*}{charginos} 
& $\chi^{\pm}_{1}$      & 152.1   \\
& $\chi^{\pm}_{2}$      & 442.9   \\
\hline
\multirow{1}{*}{gluino} 
& $\tilde{g}$                & 1221.1   \\
\hline
\multirow{1}{*}{gravitino} 
& $\tilde{G}$                & 4.6$\times 10^{-9}$   \\
\hline
\multirow{4}{*}{Higgs bosons} 
& $\tilde{h}^{0}$            & 97.4   \\
& $\tilde{H}^{0}$            & 746.5   \\
& $\tilde{A}^{0}$            & 739.3   \\
& $\tilde{H}^{\pm}$          & 748.5   \\
\hline
\end{tabular}
\label{tab:par}
\end{center}
\end{table}

\section{Signal MC production and decay}\label{sec::mc}
The mass spectrum and branching fractions were calculated using ISASUGRA 7.75 and ISASUSY 7.75~\cite{baer_b}. ISASUGRA takes input parameters from a particular  SUSY model. We used the following mGMSB parameters: \mbox{$\Lambda$ = 80 \TeV},\ \mbox{M = 240 \TeV},\ \mbox{$N_{5}$ = 2},\ \mbox{$\tan(\beta)$ = 3},\ \mbox{$sgn(\mu)$ = 1},\ \mbox{$C_{grav}$ = 1} to produce the corresponding weak scale MSSM parameter set. One of these parameters, $\mu$, was then changed accordingly to the model we used: $\mu = 0.75M_{1}$. These parameters were then taken by ISASUSY to evaluate the sparticle masses, decay rates, and branching fractions that serve as input for the program which generates the full SUSY event. About 50k events of \xxzggg, where $\ell = \mu/e$, were generated using HERWIG/JIMMY~\cite{herwig,jimmi}, which generates the sparticle cascade decays, parton showers, hadronisations, and underlying events. The full ATLAS detector simulation was done using GEANT4~\cite{geant}. All signal and background MC events were reconstructed using standard ATLAS reconstruction software. The geometrical acceptance cuts \mbox{($|\eta(\ell)|<$ 2.5)} reduced the production cross-section of \xxzggg \ to about 41 fb.

\section{Background}
The major background processes are summarized in Table~\ref{tab:bk}. We consider three types of background: (1) real $Z\gamma$ pairs from SM \zllX processes which include initial state radiation (ISR) of photons from the colliding quarks and final state photon radiation (FSR) from the $Z$ decay leptons; (2) events with one real $Z$ and jets which are misidentified as $\gamma$'s; and (3) top pairs which decay to leptons: $t\bar{t} \ra \ell + X$.

\begin{table}[!ht]
\begin{center}
\caption{Background NLO production cross-section after event selection ($\sigma_{NLO}$), fully simulated number of MC events ($N_{MC}$), and the luminosity which corresponds to the MC events. The last column indicates the MC generators used to produce the MC events.}
\begin{tabular}{|l|c|c|c|c|}
\hline
Process & \begin{tabular}{c} $\sigma_{NLO}$ \\ (pb) \end{tabular} & 
$N_{MC}$ & \begin{tabular}{c} L \\(fb$^{-1}$) \end{tabular} & Generator \\
\hline
$Z(\ra \ell^{+}\ell^{-})\gamma$ & 7.1 & 22489& 3 & {\sc PYTHIA}\\
$t\bar{t}\ra \ell + X$           & 202.9& 1860622 & 9 & MC@NLO\\
\zmmX   & 1317.8& 4909710 & 3.7 & {\sc PYTHIA}\\
\zeeX   & 1317.8& 4766732& 3.6 & {\sc PYTHIA}\\
$W^{+}Z \ra \ell\nu ll+X$      & 264.73& 20000 &75.5& {\sc MC@NLO}\\
$W^{-}Z \ra \ell\nu ll+X$      & 155.96& 15457 &99& {\sc MC@NLO}\\
$W^{+}Z \ra qq \ell\ell+X$        & 828.52& 5000  &6& {\sc MC@NLO}\\
$W^{-}Z \ra qq \ell\ell+X$        & 488.1 & 5000  &10& {\sc MC@NLO}\\
$W^{+}Z \ra \tau\nu \ell\ell+X$   & 132.37& 19675  &149& {\sc MC@NLO}\\
$W^{-}Z \ra \tau\nu \ell\ell+X$   & 77.98 & 19719  &253& {\sc MC@NLO}\\
\hline
\end{tabular}
  \label{tab:bk}
\end{center}
\end{table}

\section{Event selection}\label{sec::selection}

\begin{figure*}[t]
\begin{center}
\begin{tabular}{cc}
\includegraphics[width=80mm]{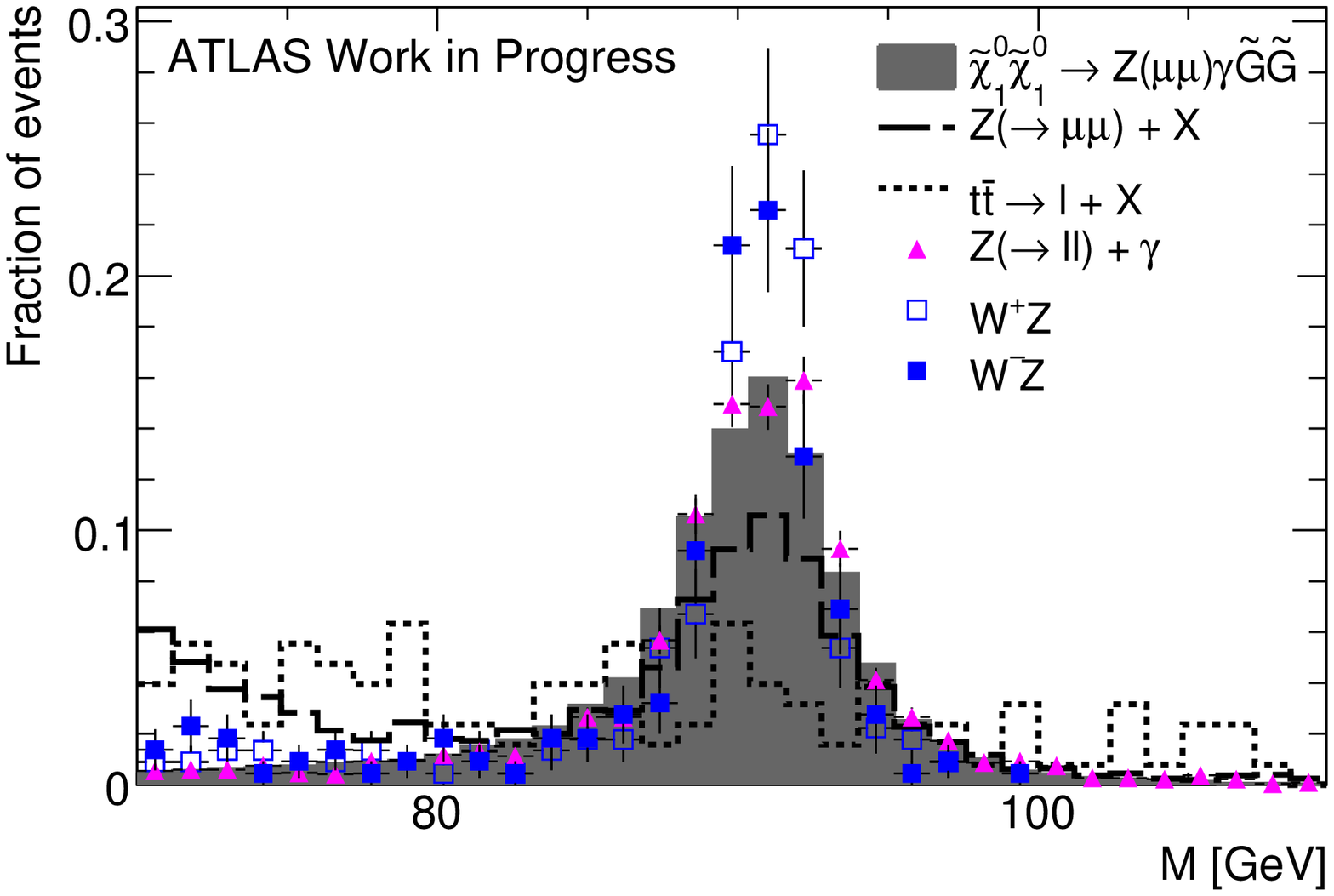} &
\includegraphics[width=80mm]{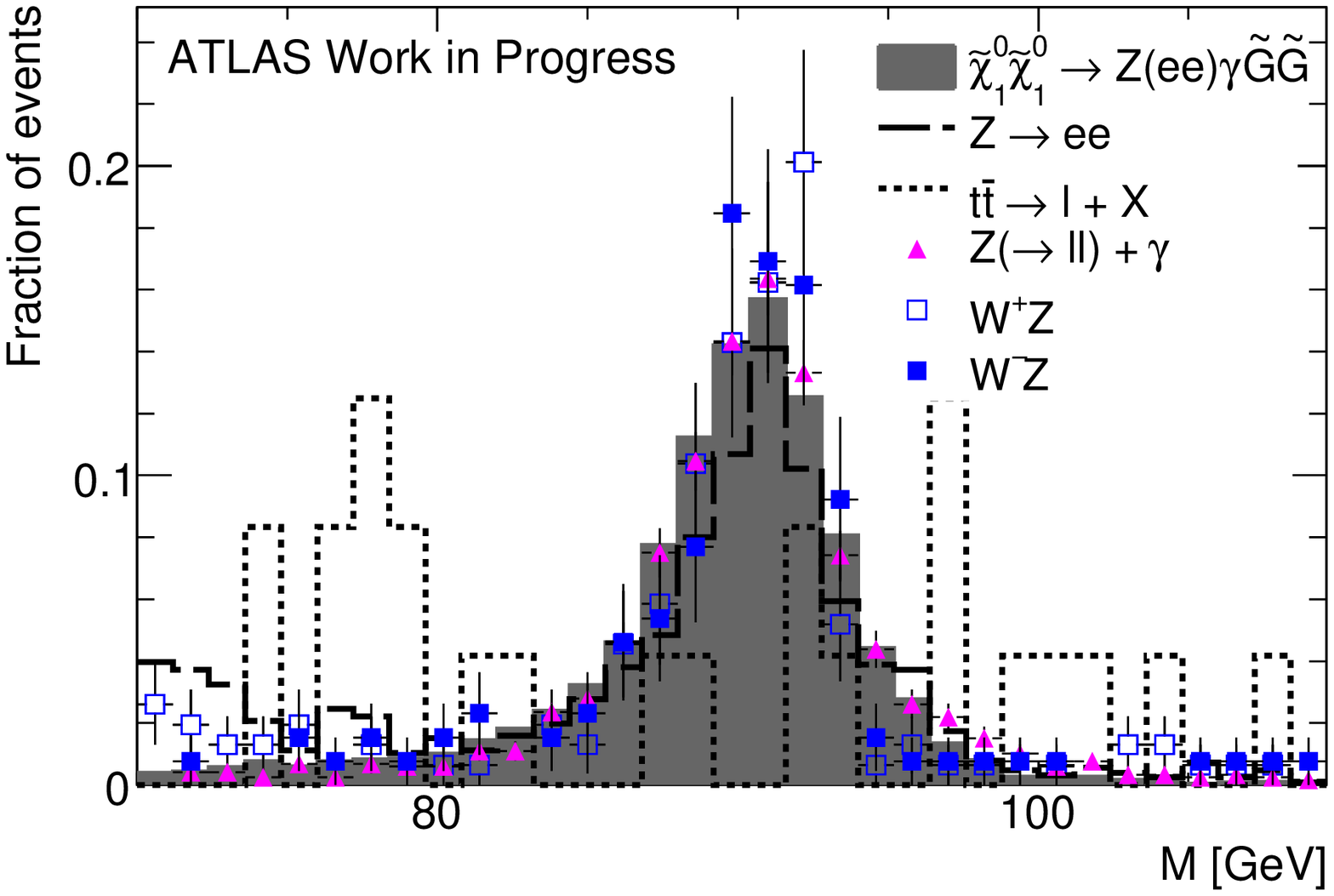} \\
\end{tabular}
\caption{Distribution of M(\mm) (left) and M(\ep) (right) after preselection. The plots are normalized to unit area.}
\label{fig:r_M}
\end{center}
\end{figure*}

To identify muons, electrons, and photons we used the following criteria:
\begin{itemize}
\item Muons are identified with an algorithm which associates a track found in the muon spectrometer, after corrections for energy loss in the calorimeter, with the corresponding inner detector track. This association is performed by statistically combining the two fitted sets of track parameters using their corresponding error matrices. The rapidity acceptance of the combined muon track is limited by the inner detector to $|\eta|<2.5$.
\item Electrons are identified using a multivariate Boosted Decision Trees (BDT) classification algorithm. BDT input variables are selected from measurements in the hadron calorimeter, the electromagnetic calorimeter and the inner detector. In the hadron calorimeter the principal variable is the fraction of hadronic $E_{T}$ leakage of the total cluster $E_{T}$. The first two longitudinal samplings of the electromagnetic calorimeter provide sets of correlated variables, including, for example, the fraction of energy deposited in each sampling. Track parameters and the number of hits in each of the inner detector layers provide a third set of variables; and finally, a set of inner detector track and EM cluster matching parameters are used. There are approximately 20 input variables incorporated by the BDT algorithm.
\item Photons are selected using the `tight' category of photon identification cuts described in~\cite{csc}.
\end{itemize}
To select events with \zll and $\gamma$, the following preselection cuts are applied. To select a good lepton candidate, the $p_{T}(\ell)$ is required to be greater than 6 \GeV with $|\eta|<2.5$. A cut of 30 \GeV is applied on the leading lepton. In the preselected lepton sample, we search for di-lepton candidates with \mbox{70 \GeV $< M (\ll) <$ 100 \GeV}. To select $\gamma$ candidates, the $p_{T}(\gamma)$ is required to be greater than 20 \GeV. To reject $\gamma$ produced through lepton bremsstrahlung radiation, the minimal value of $\Delta R$, where $\Delta R =\sqrt{\Delta\eta^{2}+\Delta\phi^{2}}$, between selected leptons and $\gamma$ is required to be greater than 0.1.
Table~\ref{tab:bk-pre} shows the efficiencies and effective remaining cross-sections after preselection cuts are applied to the signal and background processes. Figure~\ref{fig:r_M} shows the invariant mass distribution of \ll \ after preselection.

\subsection{$\mbox{\zgmet}$ selection}\label{sec:zg_sel}

\begin{table}[!ht]
\begin{center}
\caption{Results after preselection for the signal and background MC samples. The number of preselected events ($N_{ps}$), the efficiency of the preselection ($\epsilon_{ps}$) and finally the cross-section after preselection ($\sigma_{ps}$) are given.}
\begin{tabular}{|c|l|c|c|c|c|}
\hline
& Process & $N_{ps}$ & $\epsilon_{ps}$ ($\%$) &  $\sigma_{ps}$ (fb) \\
\hline
\multirow{12}{*}{2$\mu$} & \xxzmmggg    & 11574 & 49 & 10  \\
&$Z(\ra \ell^{+}\ell^{-})\gamma$ & 1845  & 8.2 & 581 \\
&$t\bar{t} \ra \ell + X$               & 127   & 0.007 & 14 \\
&\zmmX             & 2440  & 0.05  & 655 \\
&\zeeX                 & 0 & - & - \\
&$W^{+}Z \ra \ell\nu \ell\ell+X$       & 168 & 0.8 & 2 \\
&$W^{-}Z \ra \ell\nu \ell\ell+X$       & 162 & 1 & 1.6\\
&$W^{+}Z \ra qq \ell\ell+X$            & 8 & 0.16  & 1.3\\
&$W^{-}Z \ra qq \ell\ell+X$            & 8 & 0.16  & 0.8\\
&$W^{+}Z \ra \tau\nu \ell\ell+X$       & 55 & 0.28 & 0.4\\
&$W^{-}Z \ra \tau\nu \ell\ell+X$       & 55 & 0.28 & 0.2\\
\hline
\multirow{12}{*}{2$e$} & \xxzeeggg       & 6985   & 30    & 6\\
&$Z(\ra \ell^{+}\ell^{-})\gamma$  & 1187   & 5.3   & 374 \\
&$t\bar{t} \ra \ell + X$               & 25   & 0.001 & 2.7\\ 
&\zmmX             & 0    & -     & - \\
&\zeeX                 & 1263 & 0.026 & 349\\
&$W^{+}Z \ra \ell\nu \ell\ell+X$       & 116  & 0.6   & 1.5\\
&$W^{-}Z \ra \ell\nu \ell\ell+X$       & 88   & 0.6   & 0.9\\
&$W^{+}Z \ra qq \ell\ell+X$            & 9    & 0.2   & 1.5\\
&$W^{-}Z \ra qq \ell\ell+X$            & 7    & 0.14  & 0.7\\
&$W^{+}Z \ra \tau\nu \ell\ell+X$       & 38   & 0.2   & 0.25\\
&$W^{-}Z \ra \tau\nu \ell\ell+X$       & 42   & 0.2   & 0.17\\ 
\hline
\end{tabular}
\label{tab:bk-pre}
\end{center}
\end{table}

After preselection, significant background contamination remains from SM $Z(\ra \ell^{+}\ell^{-})\gamma$ and inclusive \zllX processes, which have large production cross-sections. The former has a topology similar to the signal, where a real $Z$ and $\gamma$ tend to be back-to-back in the transverse plane. However, the reconstruction efficiency of this sample is one order of magnitude lower than the signal efficiency due to the $p_{T}(\gamma)$ cut, applied in the preselection. For the latter, the source of the $\gamma$'s is mostly misidentified jets; therefore, the reconstruction efficiency is low. For further rejection of background, various distributions of kinematic variables of the background and signal MC samples were compared. 

The lepton pair is required to have an invariant mass close to the $Z$ mass, specifically $|M_{ll} - 91.2 \GeV|<13 \GeV$. This requirement is equivalent to a 5$\sigma$ width, and helps to reduce background where the lepton pair does not come directly from a $Z$.

The minimal value of $\Delta R$ between leptons and $\gamma$ is required to be greater than 0.4. This cut rejects mostly $t\bar{t}$ and \zllX events where the real $\gamma$, produced through lepton bremsstrahlung radiation, is in the same direction as leptons.

\begin{figure*}[t]
\begin{center}
\begin{tabular}{cc}
\includegraphics[width=80mm]{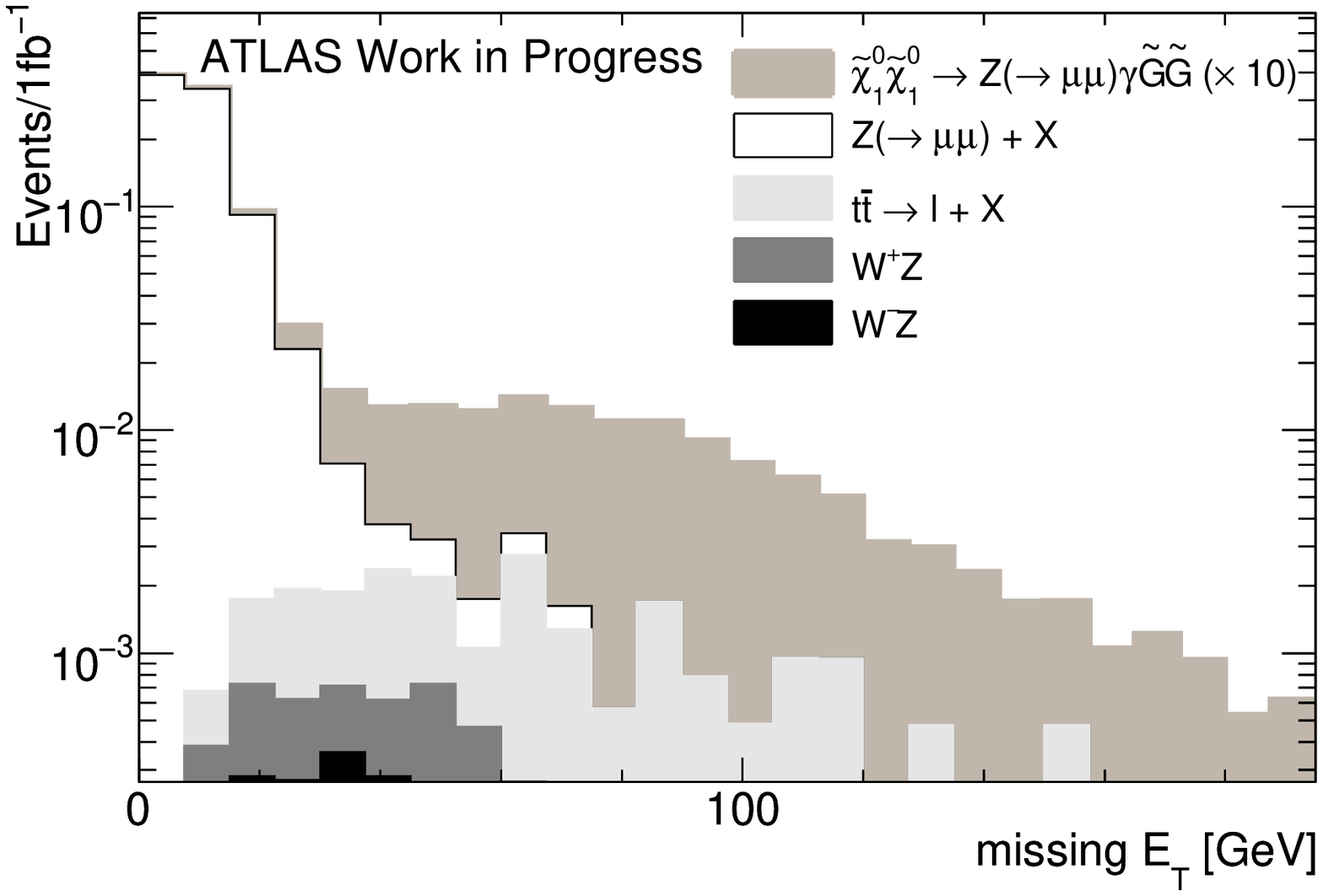} &
\includegraphics[width=80mm]{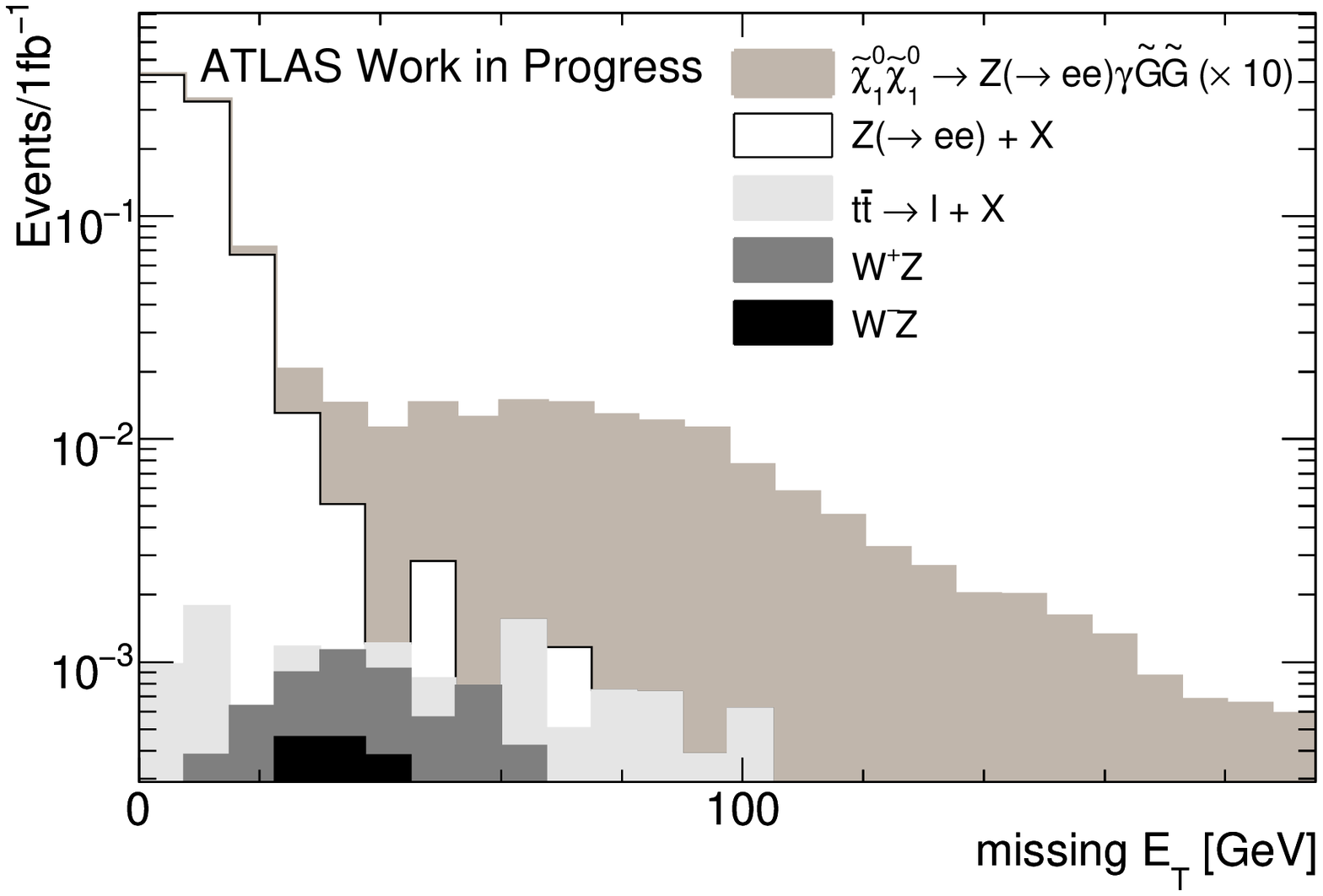} \\
\end{tabular}
\caption{Distributions of $\not\!\!\!E_{T}$ after \zmmg (left) and \zeeg (right) preselections. The plots are normalized to cross-section.}
\label{fig:r_MET}
\end{center}
\end{figure*}

Signal events are characterized by the presence of relatively large $\not\!\!E_{T}$. To reject the SM diboson processes that do not have sources of significant $\not\!\!E_{T}$, we required $\not\!\!E_{T}>40$ \GeV. Figure~\ref{fig:r_MET} shows the $\not\!\!E_{T}$ distributions of the signal and background MC events normalized to the cross-section after preselection.

The transverse like mass quantity ($M_{T}^{\textquotesingle}$)~\cite{csc} of two leptons, $\gamma$'s, and $\not\!\!E_{T}$, defined as 
\begin{equation}
M_{T}^{\textquotesingle} =
\sqrt{(p_{T}(\ell/\gamma) + \not\!\!E_{T})^{2}-\sum_{i=x,y}(\sum_{p = \ \not\!\!E_{T},\ell,\gamma}p_{i}(p))^{2}},
\label{eq:11}
\end{equation}
is highly correlated with $\not\!\!E_{T}$. Unlike the $\not\!\!E_{T}$, $M_{T}^{\textquotesingle}$ helps reject $t\bar{t}$ events. We required $M_{T}^{\textquotesingle}$ to be greater than 210 \GeV. 

The $Z\gamma$ system from the decays $\tilde{\chi}^{0}_{1} \ra \gamma \tilde{G}$ and $\tilde{\chi}^{0}_{1} \ra Z \tilde{G}$ is strongly boosted by the neutralinos and has a high invariant mass and $p_{T}$ relative to the $Z\gamma$ from the decay of \zllX. The M(Z$\gamma$) is thus required to be greater than 110 \GeV.

The photons from decays of \x \ are strongly boosted and have a high $p_{T}$ relative to the fake photons and ISR photons, which populate the low $p_{T}$ region. The $p_{T}$ of the $\gamma$ is required to be greater than 30 \GeV. The $p_{T}$ of muons and electrons from the decay $\tilde{\chi}^{0}_{1} \ra Z(\ell^+\ell^-)\tilde{G}$ is high relative to the low $p_{T}$ leptons which decay from $b$ hadrons. Therefore the cut of 10 \GeV applied to the minimum $p_{T}$ leptons helps to reject the $t\bar{t}$ background events.

In addition, $t\bar{t}$ and \zllX events can be effectively reduced by track isolation conditions on the leptons and photons. The isolation condition consists of requiring less than four tracks in a cone of $\Delta R$ = 0.3 around the 
reconstructed lepton and $\gamma$. A cut of $\Delta d_{0} < 0.1$~mm ($\Delta d_{0} < 0.15$~mm) for muons (electrons) and $\Delta z_{0} < 1$~mm, where $\Delta d_{0}$($\Delta z_{0}$) is the difference between the transverse(longitudinal) impact parameters of the tracks, helps to reject tracks from $b$ hadron decays which are expected to have, on average, large transverse and longitudinal impact parameters.

\section{Results}\label{sec::sig}
Table~\ref{tab:res} summarizes the results of the analysis for an integrated luminosity of 1 fb$^{-1}$. Since the \zllX sample includes ISR photons, the $Z(\ra \ell^{+}\ell^{-})\gamma$ sample was not included in the total background estimation to avoid the double counting of $Z\gamma$ events. After selection, four events remain in the \zmmX sample; two of these events ($0.5 \pm 0.4$ events for 1 fb$^{-1}$) include ISR photons.  This number is consistent with the $Z(\ra \ell^{+}\ell^{-})\gamma$ sample, which has one event ($0.3 \pm 0.3$ events for 1 fb$^{-1}$) that survives the selection criteria. The two remaining events surviving selection cuts from the \zmmX sample include fake photons. The number of expected events from the \zmmX sample due to jets misidentified as photons is thus expected to be $0.5 \pm 0.4$. The dominant contribution of the background comes from \zmmX (24$\%$) and $Z(\ra \ell^{+}\ell^{-})\gamma$ (24$\%$) samples after muon selection and from $t\bar{t}$ (40$\%$) and inclusive $Z$ (30$\%$) samples after electron selection.

\begin{table}[!ht]
\begin{center}
\caption{Results of cut-based analysis. This table includes the number of expected events for 1 fb$^{-1}$. The errors shown are statistical or 90$\%$ CL in the case of insufficient statistics.}
\begin{tabular}{|c|l|c|c|c|}
\hline
&  Process & \begin{tabular}{c} N of expected events \\ (for 1 fb$^{-1}$) \end{tabular} \\
\hline
\multirow{14}{*}{2$\mu$} 
& \xxzmmggg \ ($N_{s}$)           & 5.4 $\pm$ 0.06 \\
& $Z(\ra \ell^{+}\ell^{-})\gamma$ & 0.3 $\pm$ 0.3 \\
& $t\bar{t} \ra \ell + X$            & 0.3 $\pm$ 0.2 \\
& \zmmX                           & 1.1 $\pm$ 0.5 \\
& $W^{+}Z \ra \ell\nu \ell\ell+X$          & 0.28 $\pm$ 0.06 \\
& $W^{-}Z \ra \ell\nu \ell\ell+X$          & 0.26 $\pm$ 0.05 \\
& $W^{+}Z \ra qq \ell\ell+X$            & $<$ 0.4 \\
& $W^{-}Z \ra qq \ell\ell+X$            & $<$ 0.2 \\
& $W^{+}Z \ra \tau\nu \ell\ell+X$       & 0.07 $\pm$ 0.02 \\
& $W^{-}Z \ra \tau\nu \ell\ell+X$       & 0.04 $\pm$ 0.01 \\
& Total bkg ($N_{b}$) & 2.05 $\pm$ 0.6 \\
\hline
\multirow{14}{*}{2$e$} 
& \xxzeeggg ($N_{s}$)              & 3.38 $\pm$ 0.05 \\
& $Z(\ra \ell^{+}\ell^{-})\gamma$   & $<$ 0.7 \\
& $t\bar{t} \ra \ell + X$             & 0.4 $\pm$ 0.2 \\
& \zeeX            & 0.3 $\pm$ 0.3 \\
& $W^{+}Z \ra \ell\nu \ell\ell+X$           & 0.17 $\pm$ 0.05 \\
& $W^{-}Z \ra \ell\nu \ell\ell+X$           & 0.06 $\pm$ 0.02 \\
& $W^{+}Z \ra qq \ell\ell+X$             & $<$ 0.4 \\
& $W^{-}Z \ra qq \ell\ell+X$             & $<$ 0.2 \\
& $W^{+}Z \ra \tau\nu \ell\ell+X$        & 0.06 $\pm$ 0.02\\
& $W^{-}Z \ra \tau\nu \ell\ell+X$        & 0.02 $\pm$ 0.01\\
& Total bkg ($N_{b}$) & 1.0 $\pm$ 0.35  \\
\hline
\end{tabular}
\label{tab:res}
\end{center}
\end{table}

\begin{table}[!ht]
\centering
\caption{This table includes the number of expected signal/background events for 
1 fb$^{-1}$, p-value, and the significance assuming 20$\%$ systematic uncertainty, where p-value is the probability of the background to fluctuate to the total observation.}
\begin{tabular}{|l|c|c|c|c|}
\hline
            & \zmmgmet  &  \zeegmet & combined \\
\hline
$ N_{s}$ & 5.4 $\pm$ 0.06 & 3.4  $\pm$ 0.05  & 8.8 $\pm$ 0.1 \\
$ N_{b}$ & 2.05 $\pm$ 0.6 & 1.0  $\pm$ 0.35  & 3.1 $\pm$ 0.7 \\
\hline
 p-value & 0.004 &  0.04 & 0.0007\\
 Significance & 2.7$\sigma$ & 1.8$\sigma$ & 3.2$\sigma$\\
\hline
\end{tabular}
\label{tab:sig}
\end{table}
For an integrated luminosity of 1 fb$^{-1}$, 5.4 $\pm$ 0.06 and 3.4 $\pm$ 0.05 signal events ($N_{s}$) with 2.05 $\pm$ 0.6 and 1.0 $\pm$ 0.35 background events ($N_{b}$) are expected for the muon and electron samples, respectively. The detection significance, defined as the probability from a Poisson distribution with mean $N_{b}$ to observe equal or greater than $N_{s} + N_{b}$ events, converted in equivalent number of sigmas (standard deviations) of a Gaussian distribution is 2.7$\sigma$ for the 2$\mu$ sample and 1.8$\sigma$ for the 2$e$ sample, assuming 20$\%$ systematic uncertainty (Table~\ref{tab:sig}). The overall detection significance of the \xxzggg \ decay, where $\ell = \mu/e$, is expected to be 3.2$\sigma$ assuming 20$\%$ systematic uncertainty~\cite{csc} for an integrated luminosity of 1 fb$^{-1}$.
For an integrated luminosity of 3 fb$^{-1}$ the significance is expected to be 4.2$\sigma$ for muons, 3.6$\sigma$ for electrons, and 5.6$\sigma$ for the combined  samples assuming 20$\%$ systematic uncertainty. 



\section{Summary}
This note presents a study of a GMSB search at ATLAS with a $\mbox{\zgmet}$ final state where the $Z$ is reconstructed using $ee$ or $\mu\mu$ based on fully simulated MC events. The study has been performed within the GMSB model line E with $\Lambda$=80~\TeV, using 50k fully simulated signal events of \xxzggg \ that are characterized by two high $p_{T}$ leptons plus an energetic $\gamma$ with $\not\!\!E_{T}$. The results show that a significance of 5.6$\sigma$, assuming 20$\%$ systematic uncertainty, can be attained with an integrated luminosity of 3 fb$^{-1}$ in the combination of the \zeeg and \zmmg final states using a cut-based method after taking into account the background contributions from SM $t\bar{t}$, \zllX, and $WZ$. This suggests the feasibility of a search for GMSB at ATLAS with 3 fb$^{-1}$ of integrated luminosity given a center of mass energy $\sqrt{s}$ = 10 \TeV.

\bigskip 

\end{document}